\documentclass[12pt]{article}
\usepackage{amsmath, amsfonts, amssymb, amsthm}
\usepackage{graphicx}
\usepackage{here}
\begin{document}

\begin{center}
{\large\bf
Octant Degeneracy
and Plots of Parameter Degeneracy in Neutrino Oscillations Revisited}
\end{center}
\vspace*{.6cm}

\begin{center}
\large{\sc Sho Sugama and Osamu Yasuda}
\end{center}
\vspace*{0cm}
{\it
\begin{center}
Department of Physics, Tokyo Metropolitan University,\\
Hachioji, Tokyo 192-0397, Japan
\end{center}}

\vspace*{0.5cm}

{\Large \bf
\begin{center} Abstract \end{center}  }
The three kinds of parameter degeneracy
in neutrino oscillation,
the intrinsic, sign and octant degeneracy,
form an eight-fold degeneracy.
The nature of this eight-fold degeneracy
can be visualized
on the ($\sin^22\theta_{13}$, $1/\sin^2\theta_{23}$)-plane,
through quadratic curves defined by
$P(\nu_\mu\to\nu_e)=$ const. and
$P(\bar{\nu}_\mu\to\bar{\nu}_e)=$ const.,
along with a straight line $P(\nu_\mu\to\nu_\mu)=$ const.
After $\theta_{13}$ was determined by reactor
neutrino experiments, the intrinsic degeneracy
in $\theta_{13}$ transforms into
an alternative octant degeneracy in $\theta_{23}$, which can
potentially be resolved by incorporating
the value of $P(\nu_\mu\to\nu_\mu)$.
In this paper, we analytically discuss
whether this octant parameter
degeneracy is resolved or persists in the future long baseline
accelerator neutrino experiments, such as
T2HK, DUNE, T2HKK and ESS$\nu$SB.
It is found that the energy spectra near the first oscillation
maximum are effective in resolving the octant degeneracy,
whereas those near the second oscillation maximum are not.
\vspace*{0.5cm}

\section{Introduction}

In the past twenty-five years, significant progress have been achieved
in measuring neutrino oscillation parameters within the standard
three-flavor framework\,\cite{ParticleDataGroup:2022pth}.  The
ultimate goal in the study of standard neutrino oscillations is to
determine the CP phase $\delta$.  It is expected that future long
baseline experiments such as
T2HK\,\cite{Hyper-KamiokandeProto-:2015xww} and
DUNE\,\cite{DUNE:2015lol} will precisely measure the appearance
probabilities $P(\nu_\mu \to \nu_e)$ and $P(\bar{\nu}_\mu \to
\bar{\nu}_e)$, enabling the determination of $\delta$.  However, an
issue arises in extracting $\delta$ from appearance probabilities due
to the so-called parameter degeneracy.  This implies that even with
precise knowledge of $P(\nu_\mu \to \nu_e)$ and $P(\bar{\nu}_\mu \to
\bar{\nu}_e)$ at a fixed energy and baseline length, these
probabilities cannot uniquely determine the values of ($\theta_{23}$,
$\theta_{13}$, $\delta$).  Since each degenerate solution gives a
different value of $\delta$, resolving the parameter degeneracy is
crucial for the precise determination of $\delta$.
In historical context, the octant degeneracy
$\pi/4-\theta_{23}\leftrightarrow\theta_{23}-\pi/4$ was first
identified in Ref.\,\cite{Fogli:1996pv}, followed by the recognition
of the intrinsic degeneracy in Ref.\,\cite{Burguet-Castell:2001ez},
and the sign degeneracy $\Delta m^2_{\mbox{\rm\scriptsize atm}}
\leftrightarrow -\Delta m^2_{\mbox{\rm\scriptsize atm}}$ in
Ref.\,\cite{Minakata:2001qm}.  Ref.\,\cite{Barger:2001yr} highlighted
the presence of these as eight-fold degeneracies.  Moreover, a
graphical representation was introduced in Ref.\,\cite{Yasuda:2004gu}
where curves derived from $P(\nu_\mu\rightarrow\nu_e)=$ const. and
$P(\bar{\nu}_\mu\rightarrow\bar{\nu}_e)=$ const. formed quadratic
patterns on the ($X\equiv\sin^22\theta_{13}$, $Y\equiv
1/s^2_{23}$)-plane, where $s_{23}\equiv\sin\theta_{23}$.
The representation in Ref.\,\cite{Yasuda:2004gu}
visually demonstrates the existence of eight degenerate solutions, as
elaborated in the subsequent section.

After the determination of $\theta_{13}$ through reactor experiments\,
\cite{DoubleChooz:2011ymz,DayaBay:2012fng,RENO:2012mkc}, the primary
focus shifts to the sign degeneracy of $\Delta
m^2_{\mbox{\rm\scriptsize atm}}$ and the octant degeneracy of
$\theta_{23}$.  As far as the sign degeneracy is concerned, nearly all
results favor the normal ordering
(NO) over the inverted ordering (IO)\,\cite{ParticleDataGroup:2022pth}.
As for the octant degeneracy,
on the other hand, uncertainty remains regarding the correct octant,
as the Superkamiokande atmospheric neutrino results seem to favor the
lower octant, while other datasets indicate a different
preference\,\cite{Esteban:2020cvm}.

The purpose of this work is to investigate the behavior
of the octant degeneracy
when combining appearance probabilities
$P(\nu_\mu\rightarrow\nu_e)=$ const. and
$P(\bar{\nu}_\mu\rightarrow\bar{\nu}_e)=$ const.
with fixed values of $X\equiv\sin^22\theta_{13}=$ const.
and $Y\equiv 1/s^2_{23}=$ const.,
utilizing the graphical representation from
Ref.\,\cite{Yasuda:2004gu} for the
T2HK\,\cite{Hyper-KamiokandeProto-:2015xww},
T2HKK\,\cite{Hyper-Kamiokande:2016srs}, DUNE\,\cite{DUNE:2015lol}, and
ESS$\nu$SB\,\cite{ESSnuSB:2023ogw} experiments.
Our results demonstrate that, by
utilizing energy bins near the first oscillation
maximum, the octant degeneracy can be resolved by
ensuring the intersection of these quadratic curves with the vertical
line $X=$ const. and the two horizontal lines
($Y=2/(1\pm\sqrt{1-\sin^22\theta_{23}})$),
under the assumption of small experimental errors.

The paper is organized as follows.
In Sect.\,\ref{degeneracy}, we provide
background information on parameter degeneracy
and introduce the plot
illustrating the eight-fold parameter degeneracy.
Sect.\,\ref{experiments} presents
the plots for the four future long baseline experiments
and discusses the distinctive characteristics
of each experiment.  Finally, Section 4
summarizes our conclusions.

\section{The plot of the eight-fold parameter degeneracy}
\label{degeneracy}

In the second-order approximation with respect to $\theta_{13}$ and
$\Delta m_{21}^2$, the appearance probabilities $P(\nu_\mu \to \nu_e,
E)$ and $P(\bar{\nu}_\mu \to \bar{\nu}_e, E)$ for a neutrino energy $E$
and a baseline length $L$ can be expressed
as\,\cite{Cervera:2000kp,Barger:2001yr}
\begin{eqnarray}
&{\ }& \hspace{-15mm}
P(\nu_\mu \to \nu_e, E)=x^2 F^2 + 2\, \mbox{\rm sign}(\Delta m_{31}^2)\, x y F g
\cos\left[\delta+\mbox{\rm sign}(\Delta m_{31}^2)\,\Delta\right]+ y^2 g^2
\label{degene1}\\
&{\ }& \hspace{-15mm}
P(\bar{\nu}_\mu \to \bar{\nu}_e, E) =
x^2 \bar F^2 + 2\, \mbox{\rm sign}(\Delta m_{31}^2)\, x y \bar F g
\cos\left[\delta-\mbox{\rm sign}(\Delta m_{31}^2)\,\Delta\right]
+ y^2 g^2 
\,,
\label{degene2}
\end{eqnarray}
where
\begin{eqnarray}
&{\ }& \hspace{-55mm}
x \equiv s_{23} \sin 2\theta_{13}\,,
\nonumber\\
&{\ }& \hspace{-55mm}
y \equiv \left|\frac{\Delta m^2_{21}}{\Delta m^2_{31}}\right|
\cos\theta_{23} \sin 2\theta_{12}\,,
\nonumber\\
&{\ }& \hspace{-55mm}
  (F, \bar{F}) \equiv \left\{
  \begin{array}{c}
    (f, \bar{f})~~\mbox{\rm for NO}\\
    (\bar{f}, f)~~\mbox{\rm for IO}\\
  \end{array}
  \right\}\,,
\nonumber\\
&{\ }& \hspace{-55mm}
\left\{
\begin{array}{c}
  f\\
  \bar{f}
\end{array}\right\}
\equiv \frac{\sin\left(\Delta\mp A L/2\right)}
{\left(1\mp A L/2\Delta\right)}\,,
\nonumber
\end{eqnarray}
\begin{eqnarray}
&{\ }& \hspace{-75mm}
g \equiv \frac{\sin\left(A L/2\right)}
{A L/2\Delta}\,,
\nonumber\\
&{\ }& \hspace{-75mm}
\Delta \equiv \frac{|\Delta m_{31}^2| L}{4E}\,,
\nonumber\\
&{\ }& \hspace{-75mm}
A \equiv \sqrt{2}G_F N_e\,,
\nonumber
\end{eqnarray}
$G_F$ stands for the Fermi coupling constant, and $N_e$ represents the
electron density in matter, which we assume to be constant throughout
this paper.

\begin{figure}[H]
\begin{center}
\hglue -8.4cm
\includegraphics[scale=0.75]{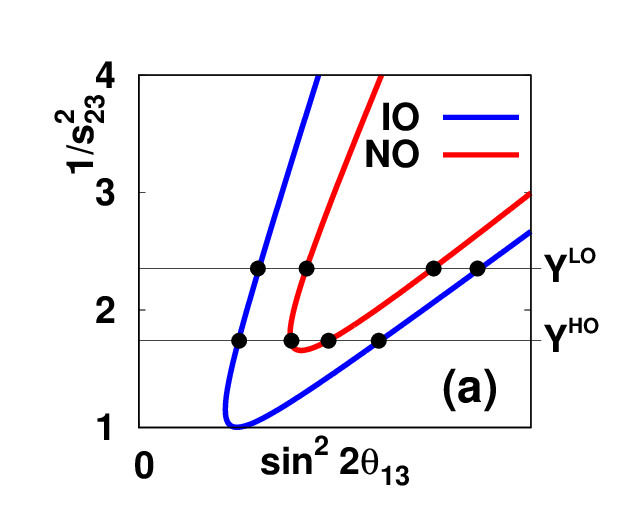}
\vglue -6.75cm
\hglue 7.4cm
\includegraphics[scale=0.75]{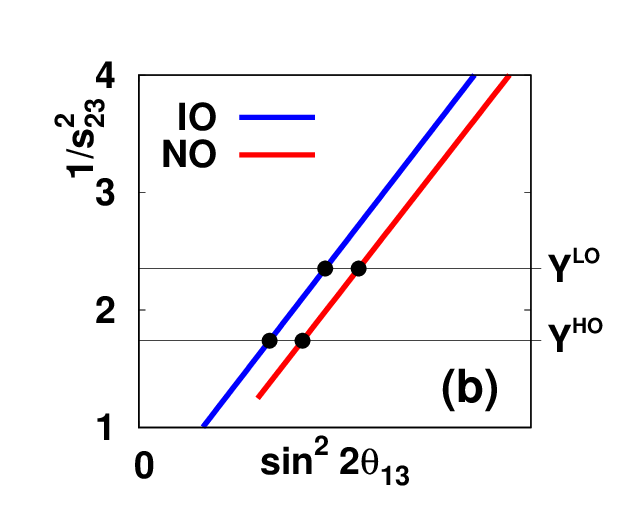}
\caption{Parameter degeneracy for normal ordering
  and inverted ordering.
  $1/s_{23}^2=Y^{\scriptsize{\mbox{\rm LO}}}$ or $1/s_{23}^2=Y^{\scriptsize{\mbox{\rm HO}}}$, where
  $Y^{\scriptsize{\mbox{\rm LO}}}$ and $Y^{\scriptsize{\mbox{\rm HO}}}$ are given by
  Eq.\,(\ref{yhlo}), is obtained from
  the disappearance probability $P(\nu_\mu\to\nu_\mu)$.
(a) Away from the oscillation maximum ($|\Delta m^2_{31}|L/4E\ne \pi/2$),~
  (b) At the oscillation maximum ($|\Delta m^2_{31}|L/4E=\pi/2$).
  In the case of (a), there are eight solutions with a different
  value of $\delta$.
  In the case of (b), four solutions exist with a two-fold degeneracy.
}
\label{fig1}
\end{center}
\end{figure}

{In the experiment, based on the measured values of the
oscillation probabilities
\begin{eqnarray}
&{\ }& \hspace{-35mm}
P(E)\equiv P(\nu_\mu \to \nu_e, E;\theta_{jk}^{\,\scriptsize{\mbox{\rm true}}},\Delta m_{jk}^{2\ \scriptsize{\mbox{\rm true}}},
\delta^{\,\scriptsize{\mbox{\rm true}}})
\label{degene5}
\end{eqnarray}
and
\begin{eqnarray}
&{\ }& \hspace{-35mm}
\bar{P}(E)\equiv P(\bar{\nu}_\mu \to \bar{\nu}_e, E;\theta_{jk}^{\,\scriptsize{\mbox{\rm true}}},
\Delta m_{jk}^{2\ \scriptsize{\mbox{\rm true}}}, \delta^{\,\scriptsize{\mbox{\rm true}}})\,,
\label{degene6}
\end{eqnarray}
which are
functions of the \textit{true} oscillation parameters and remain
fixed, we attempt to adjust the values of
$P(\nu_\mu \to \nu_e, E;\theta_{jk}^{\,\scriptsize{\mbox{\rm test}}},\Delta m_{jk}^{2\ \scriptsize{\mbox{\rm test}}},
  \delta^{\,\scriptsize{\mbox{\rm test}}})$
and $P(\bar{\nu}_\mu \to \bar{\nu}_e, E;\theta_{jk}^{\,\scriptsize{\mbox{\rm test}}},\Delta m_{jk}^{2\ \scriptsize{\mbox{\rm test}}},
  \delta^{\,\scriptsize{\mbox{\rm test}}})$, which
depend on the \textit{test} oscillation parameters, to match
$P(E)$ and $\bar{P}(E)$ by varying the \textit{test} oscillation
parameters.
The present study focuses on the following question:
Given the two appearance probabilities $P(E)$ and $\bar{P}(E)$
at a given energy $E$,
can we uniquely determine the oscillation probabilities?
In other words, do the two equations
\begin{eqnarray}
&{\ }& \hspace{-15mm}
  P(\nu_\mu \to \nu_e, E;\theta_{jk}^{\,\scriptsize{\mbox{\rm test}}},\Delta m_{jk}^{2\ \scriptsize{\mbox{\rm test}}},
  \delta^{\,\scriptsize{\mbox{\rm test}}}) = P(\nu_\mu \to \nu_e, E;\theta_{jk}^{\,\scriptsize{\mbox{\rm true}}},\Delta m_{jk}^{2\ \scriptsize{\mbox{\rm true}}},
  \delta^{\,\scriptsize{\mbox{\rm true}}})
\label{degene7}\\
&{\ }& \hspace{-15mm}
P(\bar{\nu}_\mu \to \bar{\nu}_e, E;\theta_{jk}^{\,\scriptsize{\mbox{\rm test}}},\Delta m_{jk}^{2\ \scriptsize{\mbox{\rm test}}},
  \delta^{\,\scriptsize{\mbox{\rm test}}}) =
P(\bar{\nu}_\mu \to \bar{\nu}_e, E;\theta_{jk}^{\,\scriptsize{\mbox{\rm true}}},\Delta m_{jk}^{2\ \scriptsize{\mbox{\rm true}}},
  \delta^{\,\scriptsize{\mbox{\rm true}}})
\label{degene8}
\end{eqnarray}
give unique values of the \textit{test} oscillation
parameters, in particular of the \textit{test} CP phase $\delta^{\,\scriptsize{\mbox{\rm test}}}$?
We know that the answer to this question is negative because
parameter degeneracy
affects the CP phase $\delta^{\,\scriptsize{\mbox{\rm test}}}$, as well as the mixing angles
$\theta_{13}^{\,\scriptsize{\mbox{\rm test}}}$ and $\theta_{23}^{\,\scriptsize{\mbox{\rm test}}}$, at a given neutrino energy $E$.
We will keep the values of $\Delta
m_{31}^{2\ \scriptsize{\mbox{\rm test}}}$, $\Delta m_{21}^{2\ \scriptsize{\mbox{\rm test}}}$ and $\theta_{12}^{\,\scriptsize{\mbox{\rm test}}}$ fixed,
while varying
$\delta^{\,\scriptsize{\mbox{\rm test}}}$, $\theta_{13}^{\,\scriptsize{\mbox{\rm test}}}$ and $\theta_{23}^{\,\scriptsize{\mbox{\rm test}}}$
as \textit{test} variables around
their \textit{true} values.
In the following discussions, we omit the superscript ``test''
of the \textit{test} oscillation parameters for simplicity.
In Eqs.\,(\ref{degene5}) and (\ref{degene6}),
$P(E)$ and $\bar{P}(E)$
are the probabilities evaluated using the \textit{true} oscillation parameters
$\delta^{\,\scriptsize{\mbox{\rm true}}}$,
$\theta_{13}^{\,\scriptsize{\mbox{\rm true}}}$ and $\theta_{23}^{\,\scriptsize{\mbox{\rm true}}}$ at the
 neutrino energy $E$, whereas
 $x^2 F^2 + 2\, \mbox{\rm sign}(\Delta m_{31}^2)\, x y F g
\cos\left[\delta+\mbox{\rm sign}(\Delta m_{31}^2)\,\Delta\right]+ y^2 g^2$
and $x^2 \bar F^2 + 2\, \mbox{\rm sign}(\Delta m_{31}^2)\, x y \bar F g
\cos\left[\delta-\mbox{\rm sign}(\Delta m_{31}^2)\,\Delta\right]
+ y^2 g^2 $
in Eqs.\,(\ref{degene1}) and (\ref{degene2})
are calculated with the \textit{test}
values of $\delta$, $\theta_{13}$ and $\theta_{23}$.
By introducing the new variables}
\begin{eqnarray}
&{\ }& \hspace{-84mm}
  X\equiv\sin^22\theta_{13}
\nonumber\\
&{\ }&\hspace*{-84mm}
Y\equiv \displaystyle\frac{1}{s^2_{23}}\,,
\nonumber
\end{eqnarray}
{which are functions of the \textit{test} oscillation parameters, expressing $\cos\delta$ and $\sin\delta$
from Eqs.\,(\ref{degene1}) and (\ref{degene2}) as
\begin{eqnarray}
  &{\ }&\hspace*{-15mm}
  \cos\delta = \frac{1}{4\mbox{\rm sign}(\Delta m_{31}^2)xyg\cos\Delta}
  \left(\frac{P(E)-x^2 F^2-y^2 g^2}{F}
+\frac{\bar{P}(E)-x^2 \bar F^2-y^2 g^2}{\bar{F}}\right)
\label{cosdelta}\\
&{\ }&\hspace*{-15mm}
\sin\delta = \frac{1}{4\mbox{\rm sign}(\Delta m_{31}^2)xyg\sin\Delta}
  \left(-\frac{P(E)-x^2 F^2-y^2 g^2}{F}
+\frac{\bar{P}(E)-x^2 \bar F^2-y^2 g^2}{\bar{F}}\right),
\label{sindelta}
\end{eqnarray}
}
and subsequently eliminating the \textit{test} CP phase $\delta$,
we obtain the following
expression from Eqs.\,(\ref{degene7}), (\ref{degene8}),
\,(\ref{degene1}) and (\ref{degene2})\,\cite{Yasuda:2004gu}:
\begin{eqnarray}
&{\ }&\hspace*{-10mm}
16 C X (Y-1)
\nonumber\\
&{\ }&\hspace*{-14mm}
=\frac{1}{\cos^2\Delta}\left[\left(\frac{P(E)-C}{F}
+\frac{\bar{P}(E)-C}{\bar{F}}\right)(Y-1)-(F+\bar{F})X
+\frac{P(E)}{F}+\frac{\bar{P}(E)}{\bar{F}}\right]^2\nonumber\\
&{\ }&\hspace*{-12mm}
+\frac{1}{\sin^2\Delta}\left[\left(\frac{P(E)-C}{F}
-\frac{\bar{P}(E)-C}{\bar{F}}\right)(Y-1)-(F-\bar{F})X
+\frac{P(E)}{F}-\frac{\bar{P}(E)}{\bar{F}}\right]^2,
\label{degene3}
\end{eqnarray}
where
\begin{eqnarray}
&{\ }& \hspace{-30mm}
C\equiv\left(\frac{\Delta m^2_{21}}{\Delta m^2_{31}}\right)^2
\left[\frac{\sin(A L/2)}{A L/2\Delta}\right]^2
\sin^22\theta_{12}\,.
\nonumber
\end{eqnarray}
In energy ranges where
$|\Delta m_{31}^2| L / 4E \ne n\pi/2$
($n=1,2,3,\cdots$) is satisfied,
Eq.\,(\ref{degene3}) represents a quadratic curve
in the $(X, Y)$-plane\,\cite{Yasuda:2004gu}.
{Notice that $X$ and $Y$ are the \textit{test} variables
for fitting, whereas $P(E)$ and $\bar{P}(E)$ are fixed
for a given energy $E$ because they depend on
the \textit{true} oscillation parameters and
the neutrino energy $E$.
Note also that a different point on the quadratic curve corresponds to
a different value of the \textit{test} CP phase $\delta$,
because Eqs.\,(\ref{cosdelta}) and (\ref{sindelta})
shows that $\delta$ can be uniquely defined by
$x=\sqrt{X/Y}$ and $y=|\Delta m^2_{21}/\Delta m^2_{31}|
\sin2\theta_{12}\sqrt{1-1/Y}$.}
The rationale behind choosing the variable
$Y\equiv 1/s^2_{23}$ is its analytical convenience
for demonstrating that there exist two intersections
between the curve (\ref{degene3}) and
the line $\theta_{23}=$ const.
Depending on whether the test
mass ordering is correct or wrong,
i.e., whether the test
mass ordering matches the true ordering or not,
two potential quadratic curves emerge,
as depicted in Fig.\,\ref{fig1} (a).
By utilizing the disappearance probability
$P(\nu_\mu\to\nu_\mu)\sim \sin^22\theta_{23}\sin^2(\Delta m_{31}^2L/4E)$,
we can infer the value of $\sin^22\theta_{23}$, which
yields two possibilities for{
\begin{eqnarray}
&{\ }& \hspace{-30mm}
  Y=
  \left\{
  \begin{array}{l}
    Y^{\scriptsize{\mbox{\rm HO}}}\\
    Y^{\scriptsize{\mbox{\rm LO}}}\end{array}
  \right\}\equiv \frac{2}{1\pm\sqrt{1-\sin^22\theta_{23}^{\,\scriptsize{\mbox{\rm true}}}}}\,,
\label{yhlo}
\end{eqnarray}
where the superscript HO (LO) stands for high (low) octant.
The ambiguity of choosing between $Y^{\scriptsize{\mbox{\rm HO}}}$ and
$Y^{\scriptsize{\mbox{\rm LO}}}$ represents the octant degeneracy.}
Prior to the measurement of $\theta_{13}$, therefore,
there were potentially eight solutions, as illustrated in
Fig.\,\ref{fig1} (a).
Since a different point on the quadratic curves corresponds
to a different value of the \textit{test} CP phase $\delta$,
Fig.\,\ref{fig1} (a) implies that the eight-fold degeneracy
in general
gives eight different values for the \textit{test} CP phase $\delta$.

Conversely, at the oscillation maximum
($|\Delta m_{31}^2| L / 4E =\left(n+1/2\right)\pi$;
$n=0,1,2,\cdots$),
the numerator of the first term on the right hand side
of Eq.\,(\ref{degene3}) must vanish, resulting in
a straight line in the $(X, Y)$-plane:
\begin{eqnarray}
&{\ }& \hspace{-10mm}
\left(\frac{P(E)-C}{F}
+\frac{\bar{P}(E)-C}{\bar{F}}\right)(Y-1)-(F+\bar{F})X
+\frac{P(E)}{F}+\frac{\bar{P}(E)}{\bar{F}} = 0\,.
\label{degene4}
\end{eqnarray}
In this case, two potential straight lines also emerge,
and they are illustrated in Fig.\,\ref{fig1} (b).
When considering the situation at the oscillation maximum,
there exist four feasible solutions, each
characterized by a two-fold degeneracy, as the two-fold
intrinsic degeneracy is exact in this case.

Given our primary focus on the intersection between curve
(\ref{degene3}) and $\sin^22\theta_{13}=$ const. or $1/s^2_{23}=$
const., the classification of the quadratic curve (\ref{degene3}) as
hyperbolic or elliptic does not directly influence the parameter
degeneracy characteristics.  However, it is feasible to ascertain
whether curve (\ref{degene3}) is hyperbolic or elliptic based on the
sign of the following discriminant:
\begin{eqnarray}
&{\ }& \hspace{-10mm}
  D\equiv \left[\frac{1}{\cos^2\Delta}
  \left(\frac{P(E)-C}{F}+\frac{\bar{P}(E)-C}{\bar{F}}\right)
  (F+\bar{F})\right.
\nonumber\\
&{\ }& \hspace{-2mm}
  +\left.\frac{1}{\sin^2\Delta}
  \left(\frac{P(E)-C}{F}-\frac{\bar{P}(E)-C}{\bar{F}}\right)
  (F-\bar{F})
  +8C\right]^2
\nonumber\\
&{\ }& \hspace{-2mm}
-\left[\frac{1}{\cos^2\Delta}
  \left(\frac{P(E)-C}{F}+\frac{\bar{P}(E)-C}{\bar{F}}\right)^2
  +\frac{1}{\sin^2\Delta}
   \left(\frac{P(E)-C}{F}-\frac{\bar{P}(E)-C}{\bar{F}}\right)^2\right]
\nonumber\\
&{\ }& \hspace{-2mm}
\times\left[
  \frac{1}{\cos^2\Delta}(F+\bar{F})^2
  +\frac{1}{\sin^2\Delta}(F-\bar{F})^2
\right]\,,
\label{discriminant}
\end{eqnarray}
where $D>0$ ($D<0$) corresponds to a hyperbolic (an elliptic) curve.
In general, in the energy region near the first oscillation
maximum ($|\Delta m_{31}^2| L / 4E\sim\pi/2$),
curve (\ref{degene3}) exhibits a hyperbolic shape,
whereas it can have an elliptic form in the lower energy region,
as explained in the subsequent discussions for each future
experiment.

\section{Behaviors of the plots for the future long baseline experiments}
\label{experiments}
After $\theta_{13}$ was determined, the situation of the
eight-fold parameter degeneracy is changed.
The following
three pieces of information are available:
\begin{eqnarray}
&{\ }& \hspace{-50mm}
  \mbox{\rm (i)}
    ~\mbox{\rm The~quadratic~curve~derived~from}\nonumber\\
&{\ }& \hspace{-43mm}
P(\nu_\mu \to \nu_e) = \mbox{\rm const}. ~\&~
  P(\bar{\nu}_\mu \to \bar{\nu}_e) = \mbox{\rm const}.
\label{cond1}\\
&{\ }& \hspace{-50mm}
\mbox{\rm (ii)}
~\mbox{\rm The~straight~line~} X = \mbox{\rm const}.
\mbox{\rm ~obtained~from}\nonumber\\
&{\ }& \hspace{-43mm}
P(\bar{\nu}_e \to \bar{\nu}_e) = \mbox{\rm const}.
\label{cond2}\\
&{\ }& \hspace{-50mm}
\mbox{\rm (iii)}
~\mbox{\rm The~straight~line~} Y = \mbox{\rm const}.
\mbox{\rm ~obtained~from}\nonumber\\
&{\ }& \hspace{-42mm}
P(\nu_\mu \to \nu_\mu) = \mbox{\rm const}. ~\&~
P(\bar{\nu}_\mu \to \bar{\nu}_\mu) = \mbox{\rm const}.
\label{cond3}
\end{eqnarray}
Our strategy is first to examine the intersection of
(i) and (ii), and then to enforce condition (iii).

\begin{figure}[H]
\begin{center}
\hglue -8.4cm
\includegraphics[scale=0.75]{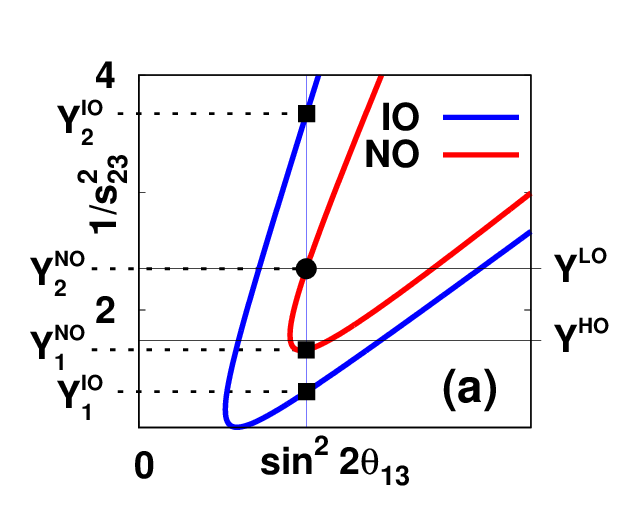}
\vglue -6.75cm
\hglue 6.9cm
\includegraphics[scale=0.75]{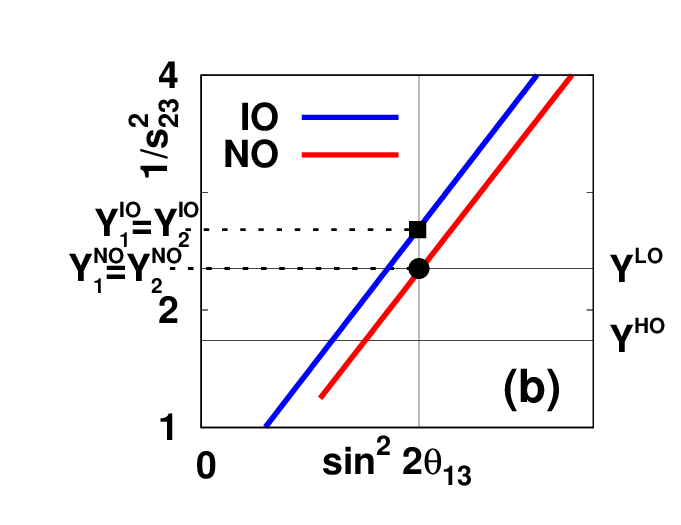}
\caption{
Parameter degeneracy after $\theta_{13}$ was determined, assuming {that
the true mass ordering is normal and the true $\theta_{23}$ lies
in the lower octant.}
The red (blue) curve corresponds to the
case where the test mass ordering is normal (inverted).
{The true solution is depicted as a black
circle, whereas the fake solutions are marked by a black square.}
In case (a),
away from the oscillation maximum, the combination of the quadratic
curves through the appearance probabilities ((i) in Eq.\,(\ref{cond1}))
and the line $X=\sin^22\theta_{13}=$ const. ((ii) in
Eq.\,(\ref{cond2})) results in four solutions {
$Y_1^{\scriptsize{\mbox{\rm NO}}}$, $Y_2^{\scriptsize{\mbox{\rm NO}}}$, $Y_1^{\scriptsize{\mbox{\rm IO}}}$, $Y_2^{\scriptsize{\mbox{\rm IO}}}$}.
In case (b), at the oscillation maximum, the
merging of the straight lines ((iii) in Eq.\,(\ref{cond3})) using the
appearance probabilities ((i) in Eq.\,(\ref{cond1})) and
$X=\sin^22\theta_{13}=$ const. ((ii) in Eq.\,(\ref{cond2})) leads to
two (two-fold) degenerate solutions {$Y_1^{\scriptsize{\mbox{\rm MO}}}=Y_2^{\scriptsize{\mbox{\rm MO}}}$~(MO=NO, IO),
where MO stands for mass ordering}.
}
\label{fig2}
\end{center}
\end{figure}

As shown in Fig.\,\ref{fig2}, if we exclusively employ (i) from
Eq.\,(\ref{cond1}) and (ii) from Eq.\,(\ref{cond2}), then generally four
solutions{, which are denoted as $Y_1^{\scriptsize{\mbox{\rm NO}}}$, $Y_2^{\scriptsize{\mbox{\rm NO}}}$, $Y_1^{\scriptsize{\mbox{\rm IO}}}$, $Y_2^{\scriptsize{\mbox{\rm IO}}}$,}
arise.  In essence, in this case, the intrinsic degeneracy
in $\theta_{13}$ is converted into an additional octant degeneracy in
$\theta_{23}$, as illustrated in Fig.\,\ref{fig2}.
{
As a strategy in this paper, we aim to determine which of the four
solutions, $Y_1^{\scriptsize{\mbox{\rm NO}}}$, $Y_2^{\scriptsize{\mbox{\rm NO}}}$, $Y_1^{\scriptsize{\mbox{\rm IO}}}$, $Y_2^{\scriptsize{\mbox{\rm IO}}}$,
matches with either $Y^{\scriptsize{\mbox{\rm LO}}}$ or
$Y^{\scriptsize{\mbox{\rm HO}}}$, thereby resolving the octant
degeneracy for a given energy.
}

In the energy region away from the oscillation maximum, as illustrated
in Fig.\,\ref{fig2} (a){, there are four intersections,
$(X,Y)=(\sin^22\theta_{13}^{\,\scriptsize{\mbox{\rm true}}},Y_j^{\scriptsize{\mbox{\rm MO}}})~(j=1,2$; MO=NO, IO), between the
$X=\sin^22\theta_{13}^{\,\scriptsize{\mbox{\rm true}}}$ and the two quadratic curves.
Namely, by imposing the conditions (i) in Eq.\,(\ref{cond1})
and (ii) in Eq.\,(\ref{cond2}), the possible solutions for
$1/s_{23}^2$ are $Y_1^{\scriptsize{\mbox{\rm NO}}}$, $Y_2^{\scriptsize{\mbox{\rm NO}}}$, $Y_1^{\scriptsize{\mbox{\rm IO}}}$, $Y_2^{\scriptsize{\mbox{\rm IO}}}$.
Furthermore, by combining the
condition (iii) in Eq.\,(\ref{cond3}) which implies that
$1/s_{23}^2$ is either $Y^{\scriptsize{\mbox{\rm LO}}}$ or $Y^{\scriptsize{\mbox{\rm HO}}}$ in Fig.\,\ref{fig2} (a),
we can reject the solutions $1/s_{23}^2= Y_1^{\scriptsize{\mbox{\rm IO}}}$ and
$1/s_{23}^2= Y_2^{\scriptsize{\mbox{\rm IO}}}$ in a particular case of Fig.\,\ref{fig2} (a).
This corresponds to the case in which the baseline
is sufficiently long to
differentiate between the mass orderings, and we can
eliminate the wrong mass ordering, which is represented by the blue
curve in Fig.\,\ref{fig2} (a).  However, since
the solution $1/s_{23}^2= Y_1^{\scriptsize{\mbox{\rm NO}}}$ is close to
$1/s_{23}^2= Y^{\scriptsize{\mbox{\rm HO}}}$ while $1/s_{23}^2= Y_2^{\scriptsize{\mbox{\rm NO}}}$ is close to
$1/s_{23}^2= Y^{\scriptsize{\mbox{\rm LO}}}$, the octant degeneracy cannot be solved
in this particular example in Fig.\,\ref{fig2} (a).}

In the energy region at the oscillation maximum, on the other hand, as
depicted in Fig.\,\ref{fig2} (b),
{there are two (two-fold degenerate) intersections,
$(X,Y)=(\sin^22\theta_{13}^{\,\scriptsize{\mbox{\rm true}}}, Y_1^{\scriptsize{\mbox{\rm MO}}}=Y_2^{\scriptsize{\mbox{\rm MO}}})$~(MO=NO, IO), between the
$X=\sin^22\theta_{13}^{\,\scriptsize{\mbox{\rm true}}}$ and the two straight lines.
In this case, the two conditions
(i) in Eq.\,(\ref{cond1}) and (ii) in Eq.\,(\ref{cond2}) leads
the two possible solutions $1/s_{23}^2 = Y_j^{\scriptsize{\mbox{\rm MO}}}~(j=1,2$; MO=NO, IO).
If these two solutions $1/s_{23}^2 = Y_j^{\scriptsize{\mbox{\rm MO}}}~(j=1,2$; MO=NO, IO)
are away from the higher octant solution
$1/s_{23}^2 = Y^{\scriptsize{\mbox{\rm HO}}}$, then the higher octant solution is
rejected and the octant degeneracy is solved.
If the two straight lines generated by the appearance
probabilities are too close, i.e., if $Y_j^{\scriptsize{\mbox{\rm IO}}}\simeq Y_j^{\scriptsize{\mbox{\rm NO}}}$,
then} it is difficult to distinguish the
true and wrong mass orderings.  It is the case with T2HK and
ESS$\nu$SB, both having relatively shorter baseline lengths.

Fig.\,\ref{fig2} is given for a specific value of the neutrino energy
and a specific baseline length{, and the values of
$Y_1^{\scriptsize{\mbox{\rm NO}}}$, $Y_2^{\scriptsize{\mbox{\rm NO}}}$, $Y_1^{\scriptsize{\mbox{\rm IO}}}$, $Y_2^{\scriptsize{\mbox{\rm IO}}}$ depend on the neutrino energy $E$.
In fact each experiment provides information of
$Y_1^{\scriptsize{\mbox{\rm NO}}}$,
$Y_2^{\scriptsize{\mbox{\rm NO}}}$,
$Y_1^{\scriptsize{\mbox{\rm IO}}}$,
$Y_2^{\scriptsize{\mbox{\rm IO}}}$ for all the energies
from its energy spectrum, and in principle
these pieces of information can be
combined to find out the correct solution.
For that purpose, we need to know
the behavior of
$Y_1^{\scriptsize{\mbox{\rm NO}}}$, $Y_2^{\scriptsize{\mbox{\rm NO}}}$,
$Y_1^{\scriptsize{\mbox{\rm IO}}}$, $Y_2^{\scriptsize{\mbox{\rm IO}}}$
across all energies in the spectrum for each experiment.
}
To observe how this degeneracy is
solved or persists {using the enerygy spectrum},
we plot the values of
$Y_1^{\scriptsize{\mbox{\rm NO}}}$, $Y_2^{\scriptsize{\mbox{\rm NO}}}$, $Y_1^{\scriptsize{\mbox{\rm IO}}}$, $Y_2^{\scriptsize{\mbox{\rm IO}}}$ as a function
of neutrino energy for the future long baseline experiments: T2HK,
DUNE, T2HKK and ESS$\nu$SB.

To generate these curves, we adopt the following
reference values, which correspond to the arithmetic
average of the best fit points
in Ref.\,\cite{ParticleDataGroup:2022pth} from the three
groups, except $\delta$ which we assume a
value of $-90^\circ$:\\
\begin{table}[H]
\begin{tabular}{|c| c| c| c|  c|  c|  c |c | }
\hline
MO
&  $\Delta m_{21}^{2\ \scriptsize{\mbox{\rm true}}}/10^{-5} \mbox{\rm eV}^2$
&  $|\Delta m_{32}^{2\ \scriptsize{\mbox{\rm true}}}|/10^{-3} \mbox{\rm eV}^2$
&  $\theta_{12}^{\,\scriptsize{\mbox{\rm true}}}$
&  $\theta_{13}^{\,\scriptsize{\mbox{\rm true}}}$
&  $\theta_{23}^{\,\scriptsize{\mbox{\rm true}}}$
&  $\delta^{\,\scriptsize{\mbox{\rm true}}}$  
&  $\rho/\mbox{\rm g}\cdot \mbox{\rm cm}^{-3}$   \\
\hline
NO  & 7.43 & 2.432 & $33.9^\circ$ & $8.49^\circ$ & $48.1^\circ$ & $-90^\circ$ &2.6 \\
IO  & 7.43 & 2.496 & $33.9^\circ$ & $8.55^\circ$ & $48.3^\circ$ & $-90^\circ$ &2.6  \\
\hline
\end{tabular}
\caption{
The reference values of the true oscillation parameters
utilized to generate plots in this study are provided for
both normal (NO) and inverted (IO) mass orderings (MO).
We choose a constant density of $\rho$ = 2.6 g/cm$^{3}$ for all the
experiments for the sake of simplicity.}
\label{table1}
\end{table}

\subsection{T2HK}
The experiment T2HK\,\cite{Hyper-KamiokandeProto-:2015xww}
features a baseline length of 295 km
and employs an off-axis neutrino
beam with an approximate energy of 0.6 GeV.
T2HK is conducted at energies near the first oscillation maximum.
Therefore, for the neutrino energy of 0.6 GeV in T2HK, the situation
is similar to that depicted in Fig.\,\ref{fig1} (b), where the
resolution of the octant degeneracy is expected to be achieved,
provided that the difference of the true and wrong values of
$1/s^2_{23}$, which is obtained from the disappearance oscillation
probabilities $P(\nu_\mu\to\nu_\mu)$ and
$P(\bar\nu_\mu\to\bar\nu_\mu)$, is larger than the experimental
errors.

\begin{figure}[H]
\hglue -1.9cm
\includegraphics[width=9.5cm]{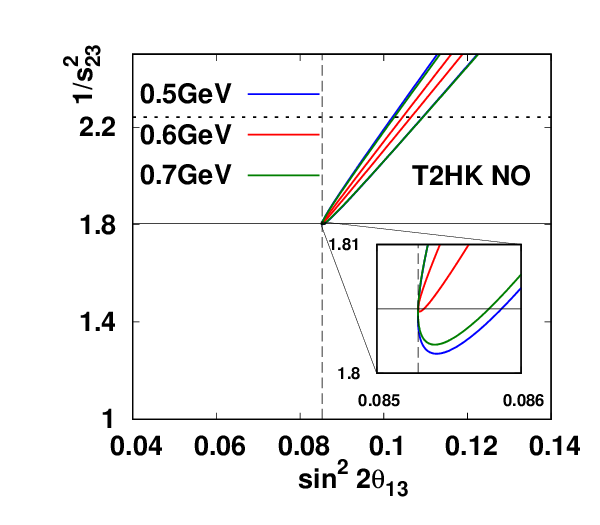}
\vglue -8.4cm
\hglue 7.4cm
\includegraphics[width=9.5cm]{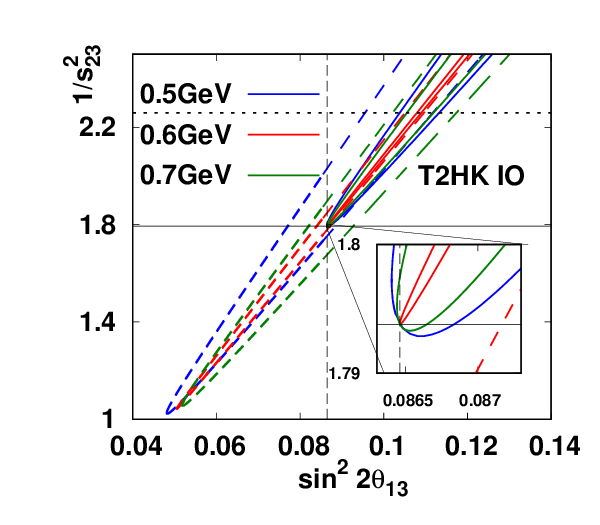}
\vglue 0.5cm
\caption{
The trajectory of the curve derived by Eq.\,(\ref{degene3}) for the
T2HK experiment is illustrated{, where a magnified view of the
region around $(X, Y) = (\sin^22\theta_{13}^{\,\scriptsize{\mbox{\rm true}}}, 1/\sin^2\theta_{23}^{\,\scriptsize{\mbox{\rm true}}})$
is supplied in each panel}.
The horizontal thin solid (dotted)
straight line represents the true (fake) value
$Y=Y^{\scriptsize{\mbox{\rm HO}}}=1/\sin^2\theta_{23}^{\,\scriptsize{\mbox{\rm true}}}$
($Y=Y^{\scriptsize{\mbox{\rm LO}}}=1/\cos^2\theta_{23}^{\,\scriptsize{\mbox{\rm true}}}$),
whereas the vertical dashed straight line indicates the value of
$X=\sin^22\theta_{13}^{\,\scriptsize{\mbox{\rm true}}}$.  The left (right) panel corresponds to the case
where the true mass ordering is normal (inverted).  The solid curves
in blue, red and green depict the trajectory of Eq.\,(\ref{degene3})
for neutrino energies of 0.5 GeV, 0.6 GeV and 0.7 GeV, respectively,
assuming the true mass ordering.  On the other hand, the dashed curves
represent trajectories assuming the wrong mass ordering.
{The energy for the dashed curves corresponds to the energy of the
solid curve of the same color.
The values of $\theta_{13}^{\,\scriptsize{\mbox{\rm true}}}$ and $\theta_{23}^{\,\scriptsize{\mbox{\rm true}}}$
depend on the true mass ordering as we can see in
Table \ref{table1}.}
}
\label{fig3}
\end{figure}

The trajectory of the curve derived from Eq.\,(\ref{degene3}) in the
($\sin^22\theta_{13}$, $1/s^2_{23}$)-plane is illustrated in
Fig.\,\ref{fig3}.
Since we assume $\delta^{\,\scriptsize{\mbox{\rm true}}}=-\pi/2$ as the true value in
this paper, if the true mass ordering is normal, then T2HK alone can
rule out the possibility of the wrong mass ordering (i.e., inverted
ordering)\,\cite{Prakash:2012az}.  Consequently, the curves
representing the wrong mass ordering, depicted as dashed curves in
Fig.\,\ref{fig3}, are absent in the left panel of Fig.\,\ref{fig3}.
{In this case, if we impose the conditions (i) in Eq.\,(\ref{cond1})
(all the solid colored curves in Fig.\,\ref{fig3})
and (ii) in Eq.\,(\ref{cond2})
(the vertical dashed line in Fig.\,\ref{fig3}),
then we are left with the
only unique solution
$(X, Y) =
(\sin^22\theta_{13}^{\,\scriptsize{\mbox{\rm true}}},
Y_1^{\scriptsize{\mbox{\rm NO}}}) = (\sin^22\theta_{13}^{\,\scriptsize{\mbox{\rm true}}},
Y_2^{\scriptsize{\mbox{\rm NO}}}) = (\sin^22\theta_{13}^{\,\scriptsize{\mbox{\rm true}}}, 1/\sin^2\theta_{23}^{\,\scriptsize{\mbox{\rm true}}})$
for the energy $E$ = 0.5, 0.6 GeV and 0.7 GeV,}
\footnote{{
Strictly speaking, the energy $E$ = 0.5, 0.7 GeV does not satisfy the
oscillation maximum condition $|\Delta m^2_{31}|L/4E=\pi/2$, so there
should be two intersections between the quadratic curve and the
vertical dashed line in the left panel of Fig.\,\ref{fig3}.  However,
from the magnified view, we see that there is approximately one
intersection between the quadratic curve and the vertical dashed line
not only for $E$ = 0.6 GeV but also for $E$ = 0.5 and 0.7 GeV.}}
{and the octant degeneracy, whether
  $Y=Y^{\scriptsize{\mbox{\rm HO}}}=1/\sin^2\theta_{23}^{\,\scriptsize{\mbox{\rm true}}}$
(the horizontal} {solid line in Fig.\,\ref{fig3}),
or $Y=Y^{\scriptsize{\mbox{\rm LO}}}=1/\cos^2\theta_{23}^{\,\scriptsize{\mbox{\rm true}}}$
(the horizontal dashed line in Fig.\,\ref{fig3}),
is resolved in the case where the true mass ordering is normal.}
Conversely, if the true mass ordering is inverted, then T2HK by itself
cannot exclude the possibility of the wrong mass ordering, leading to
the presence of the dashed curves in the right panel of
Fig.\,\ref{fig3}.
{
In this case, if we impose the conditions (i) in Eq.\,(\ref{cond1})
(all the solid-colored curves as well as all the dashed-colored curves
in Fig.\,\ref{fig3}) and (ii) in Eq.\,(\ref{cond2}) (the vertical dashed
line in Fig.\,\ref{fig3}), then we are left with three possible
solutions (for $E$ = 0.6 GeV) or four possible solutions (for
$E$ = 0.5, 0.7 GeV).  The T2HK experiment provides
information of $Y_1^{\scriptsize{\mbox{\rm NO}}}$, $Y_2^{\scriptsize{\mbox{\rm NO}}}$, $Y_1^{\scriptsize{\mbox{\rm IO}}}$, $Y_2^{\scriptsize{\mbox{\rm IO}}}$
for the energies $E$ = 0.5,  0.6 GeV, 0.7 GeV
from its energy spectrum, and in principle
we could combine them to single out the correct solution.
However, since all the values of
$Y_1^{\scriptsize{\mbox{\rm NO}}}$, $Y_2^{\scriptsize{\mbox{\rm NO}}}$, $Y_1^{\scriptsize{\mbox{\rm IO}}}$, $Y_2^{\scriptsize{\mbox{\rm IO}}}$
are approximately equal to the true value
$Y^{\scriptsize{\mbox{\rm HO}}}$, it is difficult to
pick the correct one.
Ultimately, by applying condition (iii) in Eq.\,(\ref{cond3}) (either
the horizontal solid line
{($Y=Y^{\scriptsize{\mbox{\rm LO}}}$)}
or the horizontal dotted line
{($Y=Y^{\scriptsize{\mbox{\rm HO}}}$)}
), we obtain
the unique correct solution $Y=Y^{\scriptsize{\mbox{\rm HO}}}$ for
each energy $E$.
This is because the difference between
$Y=Y^{\scriptsize{\mbox{\rm HO}}}$ and $Y=Y^{\scriptsize{\mbox{\rm LO}}}$
is larger than that between
$Y=Y_1^{\scriptsize{\mbox{\rm IO}}}$ and 
$Y=Y_j^{\scriptsize{\mbox{\rm NO}}}~(j=1,2)$ or $Y=Y_2^{\scriptsize{\mbox{\rm IO}}}$.
Therefore, the octant degeneracy is resolved even in
the case where the true mass ordering is inverted.
}

\vglue 0.5cm

\begin{figure}[H]
\hglue -1.9cm
\includegraphics[width=10.5cm]{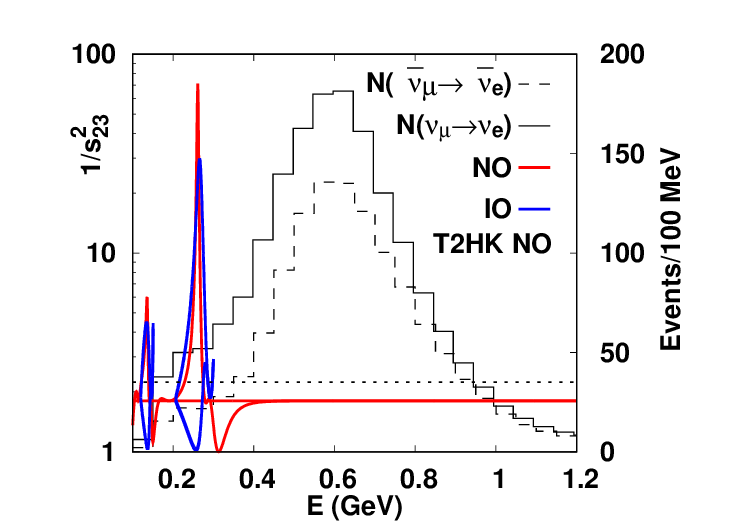}
\vglue -7.4cm
\hglue 7.5cm
\includegraphics[width=10.5cm]{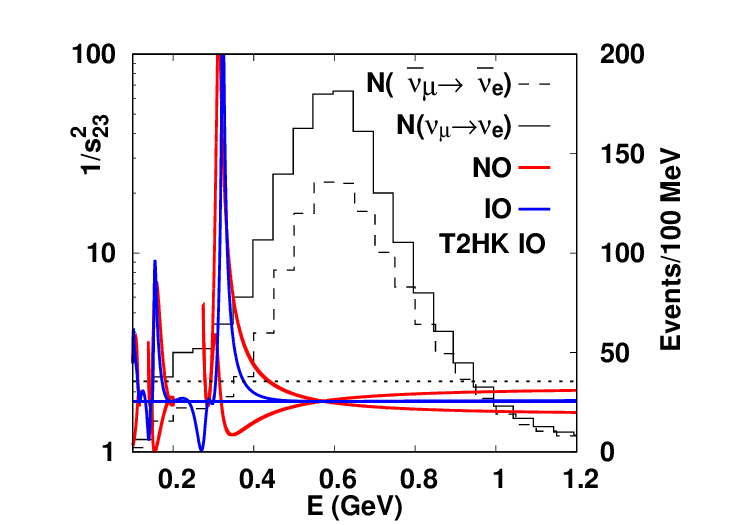}
\vglue 0.5cm
\caption{Possible values {$Y_1^{\scriptsize{\mbox{\rm NO}}}$, $Y_2^{\scriptsize{\mbox{\rm NO}}}$, $Y_1^{\scriptsize{\mbox{\rm IO}}}$, $Y_2^{\scriptsize{\mbox{\rm IO}}}$}
  for $1/s^2_{23}$ are displayed
for the case of the true mass ordering being
normal (left panel{: the two red curves represent
$Y_1^{\scriptsize{\mbox{\rm NO}}}$, $Y_2^{\scriptsize{\mbox{\rm NO}}}$, whereas the two blue curves stand for
$Y_1^{\scriptsize{\mbox{\rm IO}}}$, $Y_2^{\scriptsize{\mbox{\rm IO}}}$}) or inverted (right panel{:
the two blue curves represent
$Y_1^{\scriptsize{\mbox{\rm IO}}}$, $Y_2^{\scriptsize{\mbox{\rm IO}}}$, whereas the two red curves stand for
$Y_1^{\scriptsize{\mbox{\rm NO}}}$, $Y_2^{\scriptsize{\mbox{\rm NO}}}$}).
{There are energy ranges where the solution for
the wrong mass ordering does not exist, so
the curves for the wrong mass ordering are discontinuous.}
The horizontal dotted thin straight line
represents the fake value $Y=Y^{\scriptsize{\mbox{\rm LO}}}$ of $1/s^2_{23}${,
whereas the horizontal solid red (blue) line
in the left (right) panel stands for the true value $Y=Y^{\scriptsize{\mbox{\rm HO}}}$}.
Additionally, 
the expected numbers of events for $\nu_\mu\to\nu_e$
($\bar\nu_\mu\to\bar\nu_e$),
extracted from Ref.\,\cite{Hyper-Kamiokande:2018ofw}
under the assumption of $\delta=0$, are included for clarity,
enabling a comparison of statistical errors
across different energy regions.
}
\label{fig4}
\end{figure}

{
In general, there are four intersections between the quadratic curves
and $X=\sin^22\theta_{13}^{\,\scriptsize{\mbox{\rm true}}}$, as
depicted in the left panel of Fig.\,\ref{fig2}.  The values
$Y_1^{\scriptsize{\mbox{\rm NO}}}$, $Y_2^{\scriptsize{\mbox{\rm NO}}}$, $Y_1^{\scriptsize{\mbox{\rm IO}}}$, $Y_2^{\scriptsize{\mbox{\rm IO}}}$ of these four
intersections depend on the neutrino energy $E$.
In Fig.\,\ref{fig3} the values
$Y_1^{\scriptsize{\mbox{\rm NO}}}$, $Y_2^{\scriptsize{\mbox{\rm NO}}}$, $Y_1^{\scriptsize{\mbox{\rm IO}}}$, $Y_2^{\scriptsize{\mbox{\rm IO}}}$ are depicted
only for $E$ = 0.5 GeV, 0.6 GeV and 0.7 GeV.}
{
To observe the behavior of
$Y_1^{\scriptsize{\mbox{\rm NO}}}$, $Y_2^{\scriptsize{\mbox{\rm NO}}}$,
$Y_1^{\scriptsize{\mbox{\rm IO}}}$, $Y_2^{\scriptsize{\mbox{\rm IO}}}$
across all energies in the spectrum, Fig.\,\ref{fig4}
depicts their values against neutrino energy, spanning the entire range of the energy
spectrum.  This visualization clarifies the energy dependence of
$Y_1^{\scriptsize{\mbox{\rm NO}}}$, $Y_2^{\scriptsize{\mbox{\rm NO}}}$, $Y_1^{\scriptsize{\mbox{\rm IO}}}$, $Y_2^{\scriptsize{\mbox{\rm IO}}}$.}
They are presented for both the cases of normal ordering
(left panel{: $Y_1^{\scriptsize{\mbox{\rm NO}}}$, $Y_2^{\scriptsize{\mbox{\rm NO}}}$
  for the true mass ordering are in red,
whereas $Y_1^{\scriptsize{\mbox{\rm IO}}}$, $Y_2^{\scriptsize{\mbox{\rm IO}}}$ for the wrong mass ordering are
in blue})
and inverted ordering (right panel{:
$Y_1^{\scriptsize{\mbox{\rm IO}}}$, $Y_2^{\scriptsize{\mbox{\rm IO}}}$ for
the true mass ordering are in blue,
whereas $Y_1^{\scriptsize{\mbox{\rm NO}}}$, $Y_2^{\scriptsize{\mbox{\rm NO}}}$ for the wrong mass ordering are
in red}) as the true mass ordering.
The fake {value $Y^{\scriptsize{\mbox{\rm LO}}}$} is
represented by the horizontal dotted thin line in Fig.\,\ref{fig4}{,
whereas the true value $Y^{\scriptsize{\mbox{\rm HO}}}$
is depicted as the horizontal solid line}.
Notice that $P(E)$ and $\bar{P}(E)$ are calculated using the true
values of the oscillation parameters listed in Table \ref{table1} for
each neutrino energy $E$ and mass ordering.
In the energy range of
0.5 GeV $\lesssim E \lesssim$ 0.7 GeV, where there is a
substantial number of events for the appearance channels,
{the discrepancy between the true ($Y_1^{\scriptsize{\mbox{\rm MO}}}$)
and fake ($Y_2^{\scriptsize{\mbox{\rm MO}}}$) solutions
is small compared with that between the two octant
($Y^{\scriptsize{\mbox{\rm HO}}}$ and $Y^{\scriptsize{\mbox{\rm LO}}}$) solutions
for MO = NO, IO,
on the condition that the correct mass ordering
is assumed, i.e., 
  $|Y_2^{\scriptsize{\mbox{\rm NO}}}-Y_1^{\scriptsize{\mbox{\rm NO}}}|\ll
  |Y^{\scriptsize{\mbox{\rm HO}}}-Y^{\scriptsize{\mbox{\rm LO}}}|$
  ($|Y_2^{\scriptsize{\mbox{\rm IO}}}-Y_1^{\scriptsize{\mbox{\rm IO}}}|\ll
  |Y^{\scriptsize{\mbox{\rm HO}}}-Y^{\scriptsize{\mbox{\rm LO}}}|$)
  in the left (right) panel.
  Therefore, as in Fig.\,\ref{fig3}, where
  octant degeneracy was discussed for a given energy,
  octant degeneracy is expected to be resolved
  by taking the T2HK energy spectrum into account.}
We note in passing that, even if the true mass ordering is assumed to
be normal (the left panel of Fig.\,\ref{fig4}), solutions with wrong
mass ordering can emerge in lower energy ranges, as indicated by the
blue lines.  However, the energy in this region is
quite low, resulting in a limited number of events.

\begin{figure}[H]
\hglue 0.2cm
\includegraphics[width=15cm]{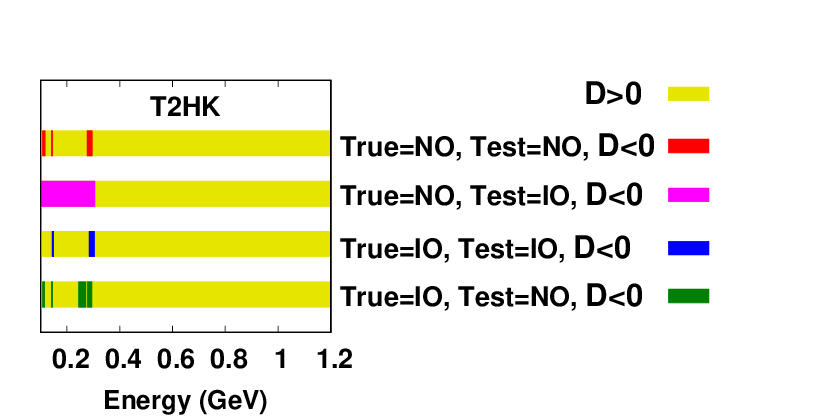}
\vglue 0.5cm
\caption{The discriminant $D$ in Eq.\,(\ref{discriminant})
for the quadratic curve (\ref{degene3})
in the case of the T2HK experiment.
}
\label{fig5}
\end{figure}

Fig.\,\ref{fig5} illustrates the sign of discriminant
(\ref{discriminant}) of the quadratic curve
(\ref{degene3}) plotted
against the neutrino energy in the case of the T2HK experiment.
In most of the energy region where 
the numbers of events of the appearance channels are large,
the discriminant is positive, indicating a hyperbolic quadratic
curve.  Conversely, and at low energies, there are regions
where the discriminant becomes negative, resulting in an
elliptic quadratic curve.

\subsection{DUNE}
The DUNE experiment\,\cite{DUNE:2015lol}
employs a baseline length of 1300 km and utilizes a wideband beam
with an average neutrino energy around 2.5 GeV.

\begin{figure}[H]
\hglue -1.9cm
\includegraphics[width=9.5cm]{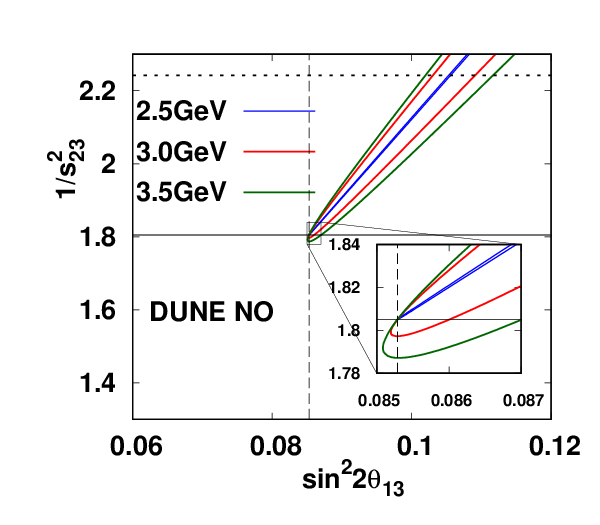}
\vglue -8.4cm
\hglue 7.2cm
\includegraphics[width=9.5cm]{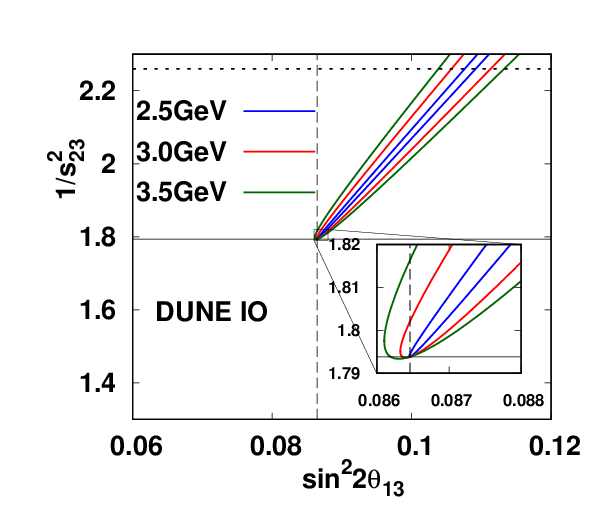}
\vglue 0.5cm
\caption{
The trajectory of the curve described by Eq.\,(\ref{degene3}) for the
DUNE experiment is depicted {with a magnified view of the
region around $(X, Y) = (\sin^22\theta_{13}^{\,\scriptsize{\mbox{\rm true}}}, 1/\sin^2\theta_{23}^{\,\scriptsize{\mbox{\rm true}}})$}.  The meaning of the straight thin lines
is the same as in Fig.\,\ref{fig3}.  The left (right) panel corresponds
to the case of true normal (inverted) mass ordering.  The solid curves
in blue, red and green represent the trajectory of
Eq.\,(\ref{degene3}) for neutrino energy of 2.5 GeV, 3.0 GeV and 3.5
GeV, respectively, assuming that the true mass ordering is normal
(inverted) in the left (right) panel.
}
\label{fig6}
\end{figure}

\begin{figure}[H]
\hglue -1.9cm
\includegraphics[width=10.5cm]{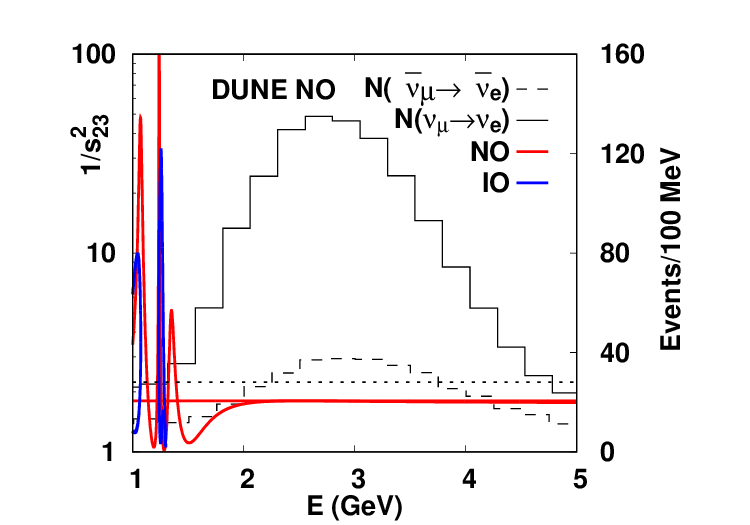}
\vglue -7.4cm
\hglue 7.5cm
\includegraphics[width=10.5cm]{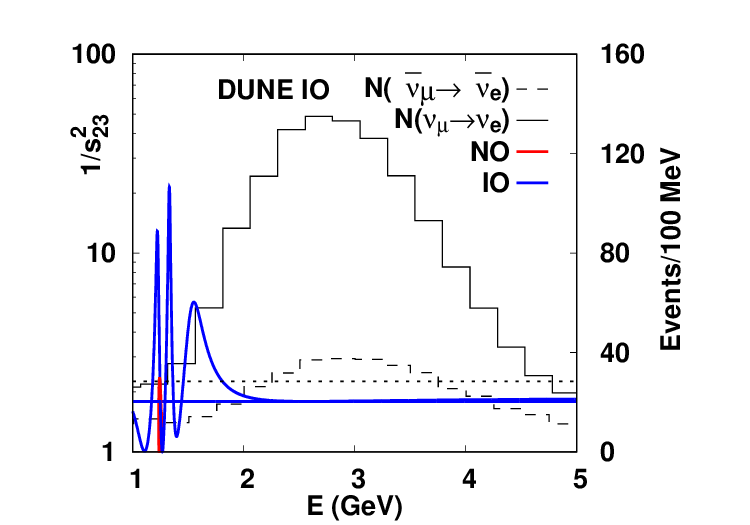}
\vglue 0.5cm
\caption{
  The values of {$Y_1^{\scriptsize{\mbox{\rm NO}}}$, $Y_2^{\scriptsize{\mbox{\rm NO}}}$, $Y_1^{\scriptsize{\mbox{\rm IO}}}$, $Y_2^{\scriptsize{\mbox{\rm IO}}}$} are
depicted for both the true normal (left panel{: the two red curves represent
$Y_1^{\scriptsize{\mbox{\rm NO}}}$, $Y_2^{\scriptsize{\mbox{\rm NO}}}$, whereas the two blue curves stand for
$Y_1^{\scriptsize{\mbox{\rm IO}}}$, $Y_2^{\scriptsize{\mbox{\rm IO}}}$}) or inverted (right
panel{:
the two red curves represent
$Y_1^{\scriptsize{\mbox{\rm IO}}}$, $Y_2^{\scriptsize{\mbox{\rm IO}}}$, whereas the two blue curves stand for
$Y_1^{\scriptsize{\mbox{\rm NO}}}$, $Y_2^{\scriptsize{\mbox{\rm NO}}}$}) mass ordering cases.
The meaning of the horizontal straight lines
is the same as in Fig.\,\ref{fig4}.
Additionally, the expected numbers of events for
$\nu_\mu\to\nu_e$ ($\bar\nu_\mu\to\bar\nu_e$), obtained from
Ref.\,\cite{DUNE:2020ypp}, are provided for reference.
}
\label{fig7}
\end{figure}

The trajectory of the curve described by Eq.\,(\ref{degene3}) for the
DUNE experiment is illustrated in Fig.\,\ref{fig6}.  Due to the
considerable baseline length, the matter effect at DUNE is
significant, which leads to the exclusion of the possibility of a
wrong mass ordering by DUNE alone, regardless of the true mass ordering.
Although DUNE employs a wideband beam, the situation for the energy
range 2.5 GeV $\lesssim E\lesssim$ 3.5 GeV is similar to that at the
first oscillation maximum.  Fig.\,\ref{fig7} presents the values of
{$Y_1^{\scriptsize{\mbox{\rm NO}}}$, $Y_2^{\scriptsize{\mbox{\rm NO}}}$, $Y_1^{\scriptsize{\mbox{\rm IO}}}$, $Y_2^{\scriptsize{\mbox{\rm IO}}}$
for both true and fake solutions as functions of neutrino
energy for the whole range of the energy spectrum},
in the case of normal ordering (left panel{: the solutions for
the true mass ordering are $Y_1^{\scriptsize{\mbox{\rm NO}}}$, $Y_2^{\scriptsize{\mbox{\rm NO}}}$,
whereas those for the wrong mass ordering are
$Y_1^{\scriptsize{\mbox{\rm IO}}}$, $Y_2^{\scriptsize{\mbox{\rm IO}}}$}) and inverted
ordering (right panel{:
the solutions for
the true mass ordering are $Y_1^{\scriptsize{\mbox{\rm IO}}}$, $Y_2^{\scriptsize{\mbox{\rm IO}}}$,
whereas those for the wrong mass ordering are
$Y_1^{\scriptsize{\mbox{\rm NO}}}$, $Y_2^{\scriptsize{\mbox{\rm NO}}}$}) as the true mass ordering.
In the energy range of
2.5 GeV $\lesssim E \lesssim$ 3.5 GeV, where
the number of events of the appearance channels is significant, 
difference between the true and fake solutions
{$Y_1^{\scriptsize{\mbox{\rm NO}}}$, $Y_2^{\scriptsize{\mbox{\rm NO}}}$, $Y_1^{\scriptsize{\mbox{\rm IO}}}$, $Y_2^{\scriptsize{\mbox{\rm IO}}}$}
is minor, suggesting a potential resolution of the octant
degeneracy, as long as
the difference of the true and wrong values {($Y^{\scriptsize{\mbox{\rm LO}}}$ and
$Y^{\scriptsize{\mbox{\rm HO}}}$)},
derived from the disappearance oscillation probabilities
((iii) in Eq.\,(\ref{cond3})),
exceeds the experimental errors.
Similar to the case of T2HK, for DUNE as well,
even if the true mass ordering is assumed to be normal, 
there can be fake solutions in lower energy ranges,
as depicted in blue
(shown in the left panel of Fig.\,\ref{fig7}),
although the energy in this region is too low to
yield a substantial number of events.  

\begin{figure}[H]
\hglue 0.2cm
\includegraphics[width=15cm]{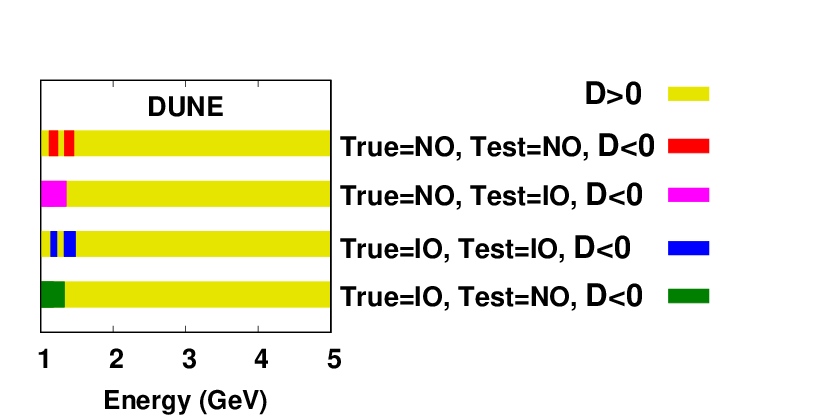}
\vglue 0.5cm
\caption{The discriminant (\ref{discriminant})
for the quadratic curve (\ref{degene3})
in the case of DUNE.
}
\label{fig8}
\end{figure}

Fig.\,\ref{fig8} displays the sign of discriminant (\ref{discriminant})
for the quadratic curve (\ref{degene3}) plotted against neutrino
energy for the DUNE case.  Similarly, in this case, the discriminant
is positive for most of the energy region where the number of events
of the appearance channels is significant.  However, at lower energies,
there are regions where the discriminant transitions to negative
values.

\subsection{T2HKK}

The T2HKK experiment\,\cite{Hyper-Kamiokande:2016srs} employs a
baseline length of 1100 km and utilizes a wideband beam with an
average neutrino energy around 1 GeV.  Unlike the T2HK and DUNE
experiments, the T2HKK experiment does not cover the energy range at
the first oscillation maximum, which occurs at $E=|\Delta
m_{31}^2|L/(4\times\pi/2)$=2.2 GeV, but rather covers the range at the
second oscillation maximum, occurring at $E=|\Delta
m_{31}^2|L/(4\times3\pi/2)$=0.75 GeV.

\begin{figure}[H]
\hglue -1.9cm
\includegraphics[width=9.5cm]{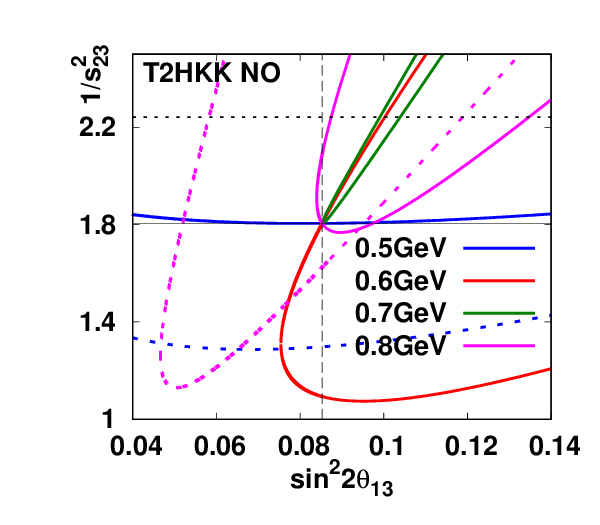}
\vglue -8.3cm
\hglue 7.2cm
\includegraphics[width=9.5cm]{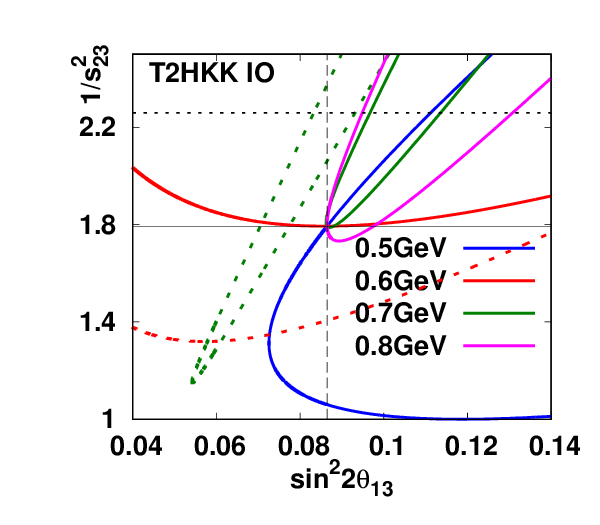}
\vglue 0.5cm
\caption{
The trajectory traced by the curve described by Eq.\,(\ref{degene3})
for the T2HKK case is showcased.  The meaning of the straight thin lines
is the same as in Fig.\,\ref{fig3}.
The left (right) panel corresponds to the true normal (inverted) mass ordering.
The solid curves colored in blue, red, green and purple denote the
trajectory of Eq.\,(\ref{degene3})
for neutrino energy of 0.5 GeV, 0.6 GeV, 0.7 GeV and 0.8 GeV,
respectively, assuming the true mass ordering
(normal (inverted) ordering in the left (right) panel).
Conversely, the dotted curves
represent trajectories assuming the wrong mass ordering.
The energy value for the dotted curves is the same as that of
the solid curve of the corresponding color.
}
\label{fig9}
\end{figure}

Fig.\,\ref{fig9} illustrates the trajectory of the curve
described by Eq.\,(\ref{degene3}).
Due to the relatively low neutrino energy, the matter effect at T2HKK
is not so large and the curves corresponding to the wrong mass ordering
appear in both true mass ordering cases.
{
Fig.\,\ref{fig9} shows that the difference between the true and fake
solutions ($Y_1^{\scriptsize{\mbox{\rm MO}}}$ and
$Y_2^{\scriptsize{\mbox{\rm MO}}}$) is large for each energy and for
each mass ordering, giving the impression that it helps to resolve the
octant degeneracy. However, the dependence of the location of these
intersections on the energy $E$ is so strong that it makes it
difficult to resolve the degeneracy, as we will see later.}

\begin{figure}[H]
\hglue -1.9cm
\includegraphics[width=10.5cm]{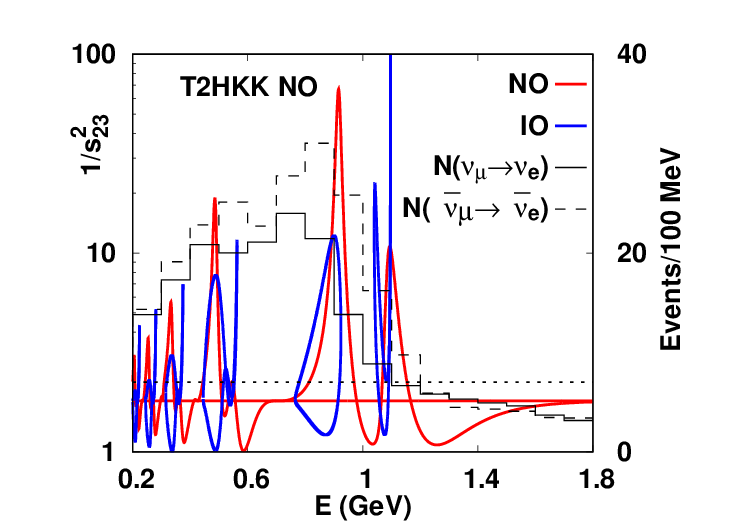}
\vglue -7.4cm
\hglue 7.5cm
\includegraphics[width=10.5cm]{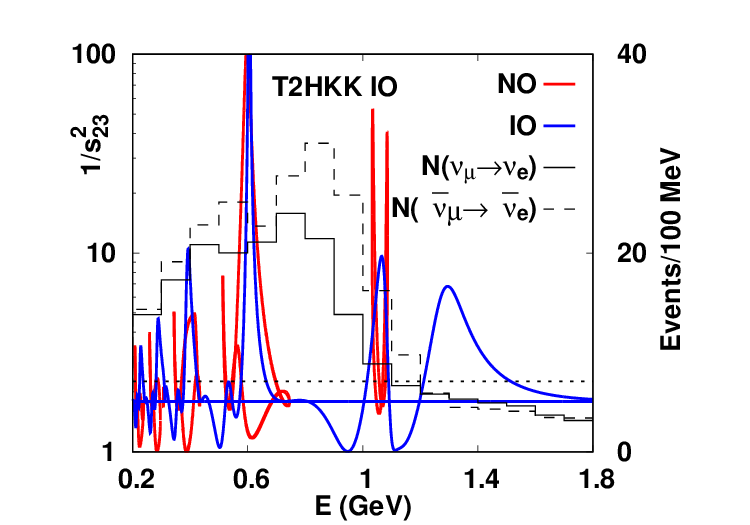}
\vglue 0.5cm
\caption{
The values {$Y_1^{\scriptsize{\mbox{\rm NO}}}$, $Y_2^{\scriptsize{\mbox{\rm NO}}}$, $Y_1^{\scriptsize{\mbox{\rm IO}}}$, $Y_2^{\scriptsize{\mbox{\rm IO}}}$} of the two solutions
for $1/s^2_{23}$ are displayed, with the left panel
representing the case of true normal mass ordering
and the right panel depicting the true inverted mass ordering case.
Similar to Fig.\,\ref{fig4}, the horizontal dotted thin line
indicates the fake value {$Y=Y^{\scriptsize{\mbox{\rm LO}}}$} of $1/s^2_{23}$ derived from
the disappearance oscillation probabilities
((iii) in Eq.\,(\ref{cond3})){, whereas the horizontal solid red (blue) line
in the left (right) panel stands for the true value $Y=Y^{\scriptsize{\mbox{\rm HO}}}$}.
The meaning of the horizontal straight lines
is the same as in Fig.\,\ref{fig4}.
For the sake of clarity and to facilitate comparison,
the expected numbers of events for $\nu_\mu\to\nu_e$
($\bar\nu_\mu\to\bar\nu_e$) are provided, sourced
from Ref.\,\cite{Hyper-Kamiokande:2016srs}.
}
\label{fig10}
\end{figure}

Fig.\,\ref{fig10} presents the values of $1/s^2_{23}$ for
both true and fake solutions as functions of neutrino energy,
for the case of normal ordering (left panel{: the solutions of
$1/s^2_{23}$ for
the true mass ordering are $Y_1^{\scriptsize{\mbox{\rm NO}}}$, $Y_2^{\scriptsize{\mbox{\rm NO}}}$,
whereas those for the wrong mass ordering are
$Y_1^{\scriptsize{\mbox{\rm IO}}}$, $Y_2^{\scriptsize{\mbox{\rm IO}}}$})
and inverted ordering (right panel{:
the solutions of $1/s^2_{23}$ for
the true mass ordering are $Y_1^{\scriptsize{\mbox{\rm IO}}}$, $Y_2^{\scriptsize{\mbox{\rm IO}}}$,
whereas those for the wrong mass ordering are
$Y_1^{\scriptsize{\mbox{\rm NO}}}$, $Y_2^{\scriptsize{\mbox{\rm NO}}}$}) as the true mass ordering.
The behavior of the curves is sensitive to the
neutrino energy in the energy range below the
first oscillation maximum (i.e., 2.2 GeV).
Even in the energy range
0.5 GeV $\lesssim E \lesssim$ 0.8 GeV, where
the number of events of the appearance channels is significant, 
the difference between the values {$Y_1^{\scriptsize{\mbox{\rm NO}}}$, $Y_2^{\scriptsize{\mbox{\rm NO}}}$, $Y_1^{\scriptsize{\mbox{\rm IO}}}$, $Y_2^{\scriptsize{\mbox{\rm IO}}}$}
of $1/s^2_{23}$
for true and fake solutions is substantial compared to the 
difference between the values of the lower {($Y^{\scriptsize{\mbox{\rm LO}}}$)}
and higher {($Y^{\scriptsize{\mbox{\rm HO}}}$)} octant
solutions derived from the disappearance channels
$\nu_\mu\to\nu_\mu$ and $\bar\nu_\mu\to\bar\nu_\mu$.
Furthermore, it is evident that the energy dependence
of the difference between the values
{$Y_1^{\scriptsize{\mbox{\rm NO}}}$, $Y_2^{\scriptsize{\mbox{\rm NO}}}$, $Y_1^{\scriptsize{\mbox{\rm IO}}}$, $Y_2^{\scriptsize{\mbox{\rm IO}}}$} of $1/s^2_{23}$
for the true and fake solutions is significant.  In other words,
a slight change in neutrino energy leads
to a considerable change in the $1/s^2_{23}$ value for both true
and fake solutions, causing it to shift from below the correct value
to above it.  Consequently, integrating this difference over a certain
neutrino energy interval yields results that depend on the size of the
energy interval, rendering a reliable conclusion difficult to
obtain.  Hence, it is not expected that the T2HKK experiment alone
will resolve the octant degeneracy.
It has been previously reported\,\cite{Panda:2022vdw} that the detector
located at 1100 km in the T2HKK experiment exhibits limited
sensitivity to octant degeneracy.  The aforementioned discussion
regarding the behavior near the second oscillation maximum provides a
rationale for the restricted sensitivity observed in T2HKK.
However, it is expected that
the T2HKK experiment will be combined with the T2HK
experiment, which is likely to result in the resolution of the octant
degeneracy.

\begin{figure}[H]
\hglue 0.2cm
\includegraphics[width=15cm]{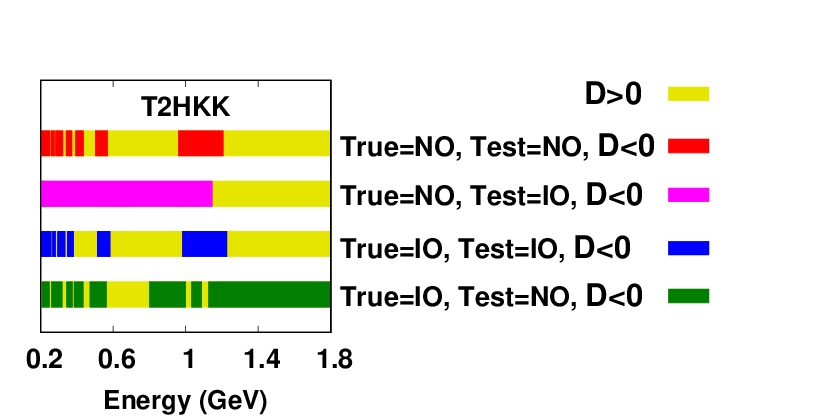}
\vglue 0.5cm
\caption{The discriminant (\ref{discriminant})
for the quadratic curve (\ref{degene3})
in the case of T2HKK.
}
\label{fig11}
\end{figure}

Fig.\,\ref{fig11} visualizes the sign of the discriminant
(\ref{discriminant}) associated with the quadratic curve
(\ref{degene3}), portrayed as a function of the neutrino energy
for the T2HKK case.
Unlike the cases of T2HK and DUNE, even in regions where 
the number of events in the appearance channels is substantial,
the discriminant takes on negative values.

\subsection{ESS$\nu$SB}

\begin{figure}[H]
\hglue -1.9cm
\includegraphics[width=9.5cm]{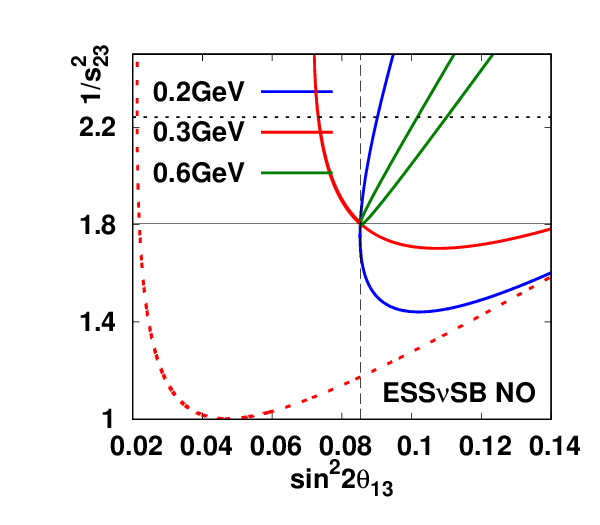}
\vglue -8.3cm
\hglue 7.2cm
\includegraphics[width=9.5cm]{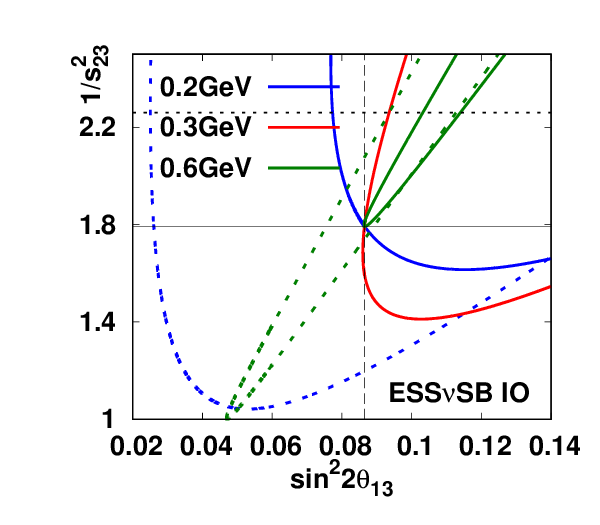}
\vglue 0.5cm
\caption{Trajectory of the curve described by Eq.\,(\ref{degene3}) for
the ESS$\nu$SB case.  The meaning of the straight thin lines is the
same as in Fig.\,\ref{fig3}.  The left (right) panel corresponds to
the case of true normal (inverted) mass ordering.  The solid curves
in blue, red and green illustrate the trajectory of
Eq.\,(\ref{degene3}) for neutrino energy of 0.2 GeV, 0.3 GeV and 0.6
GeV, respectively, assuming the true mass ordering (normal
(inverted) ordering in the left (right) panel).
}
\label{fig12}
\end{figure}
The ESS$\nu$SB experiment,\cite{ESSnuSB:2023ogw} utilizes a baseline
length of 360 km\footnote{During the earlier stages of the ESS$\nu$SB
plan, alternative baseline lengths were considered, but the baseline
length is now established at 360 km.} and employs a wideband beam with
an average neutrino energy of approximately 0.4 GeV.  The ESS$\nu$SB
experiment aims to cover energy ranges corresponding to both the first
oscillation maximum ($E\sim$ 0.73 GeV) and the second oscillation
maximum ($E\sim$ 0.24 GeV). 

Fig.\,\ref{fig12} illustrates the trajectory of the curve described by
Eq.\,(\ref{degene3}).  Due to its relatively short baseline length, the
matter effect at ESS$\nu$SB is small, and curves corresponding to the
wrong mass ordering appear in both true mass orderings.  An exception
is observed at 0.6 GeV for the true normal mass ordering, where the
wrong mass ordering is rejected for the same reason as T2HK.
{As in the case of T2HKK, Fig.\,\ref{fig12}
demonstrates that the difference between the true and fake solutions
($Y_1^{\scriptsize{\mbox{\rm MO}}}$ and $Y_2^{\scriptsize{\mbox{\rm MO}}}$)
is significant for each energy and each mass ordering.  However, the
dependence of the location of these intersections on the energy $E$ is
strong.}

\begin{figure}[H]
\hglue -1.9cm
\includegraphics[width=10.5cm]{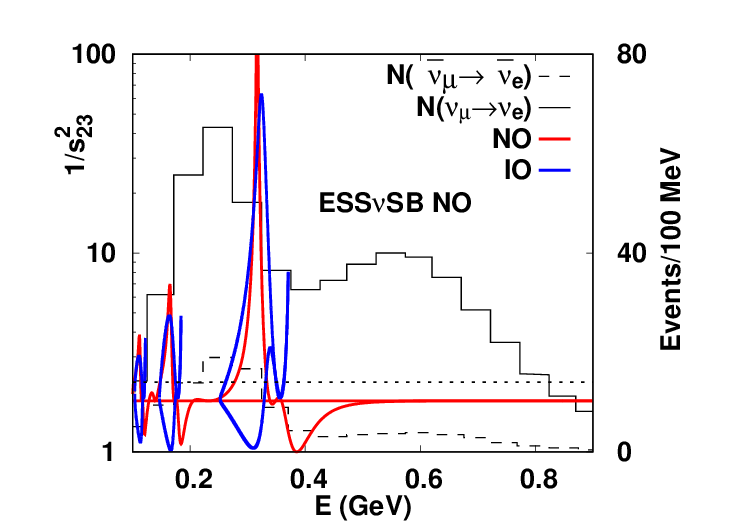}
\vglue -7.4cm
\hglue 7.5cm
\includegraphics[width=10.5cm]{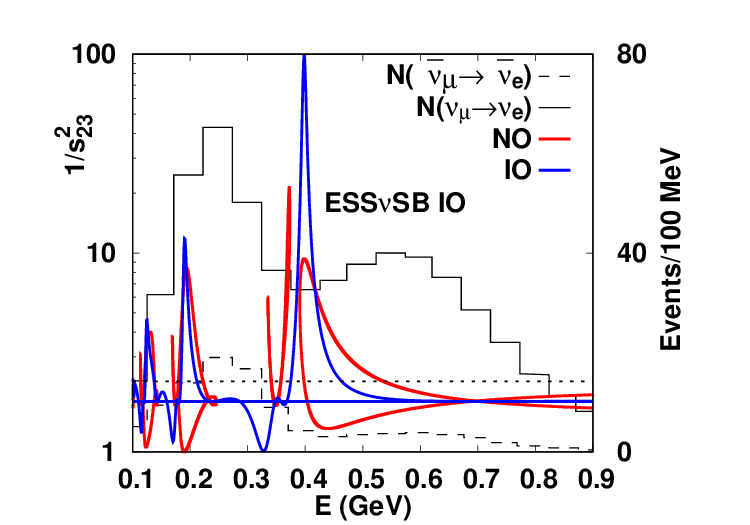}
\vglue 0.5cm
\caption{The values of {$Y_1^{\scriptsize{\mbox{\rm NO}}}$,
  $Y_2^{\scriptsize{\mbox{\rm NO}}}$, $Y_1^{\scriptsize{\mbox{\rm IO}}}$,
  $Y_2^{\scriptsize{\mbox{\rm IO}}}$}
  are shown for the case where the true mass ordering is
  normal (left panel{: the two red curves represent
  $Y_1^{\scriptsize{\mbox{\rm NO}}}$, $Y_2^{\scriptsize{\mbox{\rm NO}}}$,
  whereas the two blue curves stand for
  $Y_1^{\scriptsize{\mbox{\rm IO}}}$, $Y_2^{\scriptsize{\mbox{\rm IO}}}$})
  or inverted (right panel{:
  the two red curves represent
  $Y_1^{\scriptsize{\mbox{\rm IO}}}$, $Y_2^{\scriptsize{\mbox{\rm IO}}}$,
  whereas the two blue curves stand for
$Y_1^{\scriptsize{\mbox{\rm NO}}}$, $Y_2^{\scriptsize{\mbox{\rm NO}}}$}).  
The meaning of the horizontal straight lines
is the same as in Fig.\,\ref{fig4}.
Additionally, the expected
numbers of events for $\nu_\mu\to\nu_e$ ($\bar\nu_\mu\to\bar\nu_e$),
taken from Ref.\,\cite{Alekou:2022mav}, are provided for reference. }
\label{fig13}
\end{figure}

Fig.\,\ref{fig13} illustrates the value of $1/s^2_{23}$ for both true
and fake solutions, presented as functions of neutrino energy in the
case of normal ordering (left panel{: the solutions of
$1/s^2_{23}$ for
the true mass ordering are $Y_1^{\scriptsize{\mbox{\rm NO}}}$, $Y_2^{\scriptsize{\mbox{\rm NO}}}$,
whereas those for the wrong mass ordering are
$Y_1^{\scriptsize{\mbox{\rm IO}}}$, $Y_2^{\scriptsize{\mbox{\rm IO}}}$}) or inverted one (right panel{:
the solutions of $1/s^2_{23}$ for
the true mass ordering are $Y_1^{\scriptsize{\mbox{\rm IO}}}$, $Y_2^{\scriptsize{\mbox{\rm IO}}}$,
whereas those for the wrong mass ordering are
$Y_1^{\scriptsize{\mbox{\rm NO}}}$, $Y_2^{\scriptsize{\mbox{\rm NO}}}$}) as
the true mass ordering.  The behavior of the curves around the energy
region of the first oscillation maximum is analogous to that observed
in T2HK and DUNE.  In this energy range, resolving the octant
degeneracy becomes potentially feasible if a substantial number of
appearance events are available for both neutrinos and antineutrinos.
However, practical observations from Fig.\,\ref{fig13} reveal that the
numbers of events $N(\bar\nu_\mu\to\bar\nu_e)$ are relatively low
around the energy range corresponding to the first oscillation
maximum.  Consequently, the resulting larger experimental errors make
it challenging to effectively resolve the octant degeneracy.
Furthermore, the behavior of the curves around the energy region of the
second oscillation maximum resembles that of T2HKK.  Octant degeneracy
is not likely to be resolved solely using the energy bins around this
range.  Prior
studies\,\cite{Agarwalla:2014tpa,Chakraborty:2019jlv,ESSnuSB:2021azq}
have highlighted the limited sensitivity of ESS$\nu$SB to octant
degeneracy.  The explanations provided above regarding behaviors near
the first and second oscillation maxima clarify the reasons for the
observed limitations in ability of ESS$\nu$SB to address octant
degeneracy.

\begin{figure}[H]
\hglue 0.2cm
\includegraphics[width=15cm]{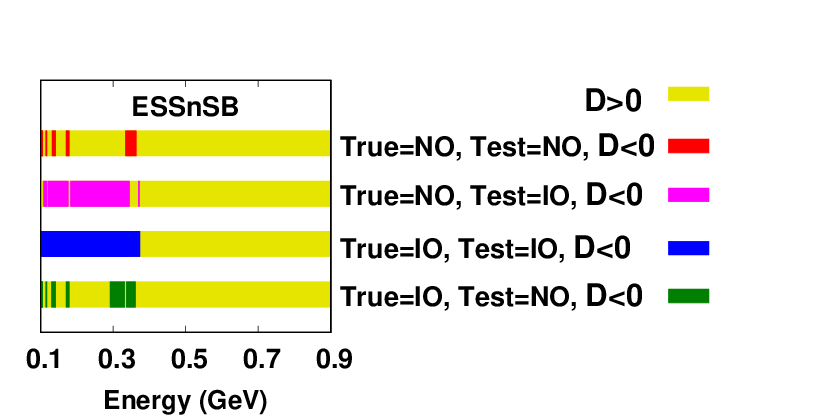}
\vglue 0.5cm
\caption{The discriminant (\ref{discriminant}) of the quadratic curve
(\ref{degene3}) in the case of ESS$\nu$SB.}
\label{fig14}
\end{figure}

Fig.\,\ref{fig14} illustrates the sign of the discriminant
(\ref{discriminant}) for the quadratic curve
(\ref{degene3}) plotted as a
function of the neutrino energy
in the case of ESS$\nu$SB.
Similar to the cases of T2HK and DUNE,
the discriminant remains positive for the energy range
around the first oscillation maximum ($E\sim$ 0.73 GeV), while it
transitions to negative values around $E\sim$ 0.3 GeV.

From these plots, a general trend emerges suggesting that octant
degeneracy can potentially be resolved using energy bins around the
first oscillation maximum, assuming sufficiently large numbers of
appearance events in this energy range for both neutrinos and
antineutrinos.  However, it is anticipated to be challenging to
effectively address the degeneracy using energy bins centered around
the second oscillation maximum.

\section{Conclusions}
\label{conclusions}
In this paper we conducted an analytical investigation of octant
degeneracy after the determination of $\theta_{13}$, utilizing the
trajectory defined by the appearance probabilities in the
($\sin^22\theta_{13}$, $1/s^2_{23}$)-plane.
By considering the appearance channels $\nu_\mu\to\nu_e$ and
$\bar{\nu}_\mu\to\bar{\nu}_e$, along with
the reactor data on $\theta_{13}$, we identified four
potential solutions for $1/s^2_{23}$.
Incorporating the disappearance channels $\nu_\mu\to\nu_\mu$ and
$\bar{\nu}_\mu\to\bar{\nu}_\mu$, it is theoretically feasible
to resolve octant degeneracy, assuming that
experimental errors are sufficiently small compared to the difference
between the true and fake solutions.

We examined the values of $1/s^2_{23}$ for the true and fake solutions
in future long baseline experiments, including T2HK, DUNE, T2HKK, and
ESS$\nu$SB.  T2HK and DUNE have adequate numbers of appearance events
for both neutrinos and antineutrinos in the energy region near the
first oscillation maximum, where the difference between the values of
$1/s^2_{23}$ for the true and fake solutions is so minor that the fake
octant solution inferred from the disappearance channels can be
excluded.  As a result, these experiments hold the potential to
effectively resolve the octant degeneracy.  On the other hand, T2HKK
and ESS$\nu$SB encounter challenges due to insufficient numbers of
appearance events for either
{antineutrinos or both neutrinos and antineutrinos}
around the
first oscillation maximum.  Moreover, the significant energy-dependent
variation between the true and fake solutions around the second
oscillation maximum in these experiments makes it difficult to solely
overcome the degeneracy through experimental data.

Additionally, we calculated the discriminant of the quadratic curves
as a function of energy, illustrating the energy regions where they
exhibit hyperbolic or elliptic behavior for each experiment.  This
aspect was not discussed in Ref.\cite{Yasuda:2004gu}.

It is important to note that our study relies on discussions centered
around the best-fit points of oscillation parameters. In practice,
however, due to experimental errors, conclusions of octant
degeneracy resolution should be interpreted with a certain level of
confidence in their potential.  Our primary aim in this study is to
provide analytical insights into which experiments and energy ranges
show promise for resolving the octant degeneracy.  We hope that our
work contributes valuable insights into how octant degeneracy might be
addressed in future long baseline experiments.

\section*{Acknowledgments}
One of the authors (O.Y.) would like to thank
Monojit Ghosh for discussions.
This research was partly supported by a
Grant-in-Aid for Scientific Research of the Ministry of Education,
Science and Culture, under Grant No. 21K03578.

\end{document}